# A transferable machine-learning framework linking interstice distribution and plastic heterogeneity in metallic glasses

Qi Wang, Anubhav Jain

Lawrence Berkeley National Laboratory, Energy Technologies Area,

1 Cyclotron Road, Berkeley, CA 94720

Email: qiwang.mse@gmail.com; ajain@lbl.gov

## ABSTRACT

When metallic glasses (MGs) are subjected to mechanical loads, the plastic response of atoms is non-uniform. However, the extent and manner in which atomic environment signatures present in the undeformed structure determine this plastic heterogeneity remain elusive. Here, we demonstrate that novel site environment features that characterize interstice distributions around atoms combined with machine learning (ML) can reliably identify plastic sites in several Cu-Zr compositions. Using only quenched structural information as input, the ML-based plastic probability estimates ("quench-in softness" metric) can identify plastic sites that could activate at high strains, losing predictive power only upon the formation of shear bands. Moreover, we reveal that a quench-in softness model trained on a single composition and quenching rate substantially improves upon previous models in generalizing to different compositions and completely different MG systems ($Ni_{62}Nb_{38}$, $Al_{90}Sm_{10}$ and $Fe_{80}P_{20}$). Our work presents a general, data-centric framework that could potentially be used to address the structural origin of any site-specific property in MGs.

# 1. Introduction

Upon sufficiently rapid cooling, many metallic melts become "frozen" and form amorphous alloys, or metallic glasses (MGs)[1–5]. The combination of "metal" and "glass" not only produces many technologically useful properties but also introduces intriguing and incompletely understood behaviors[6–8]. Understanding and controlling deformation is one of the greatest challenges in MGs[9–15]. Unlike in crystals, each atom in a disordered material has a unique atomic environment, and as a result, when subjected to mechanical stimuli, their response can, in principal, be different. This heterogeneity makes it notoriously difficult to establish a causal link between structure and deformation[9–11,14,15].

Several indicators have previously been proposed to characterize the local glass structure and serve as indicators of plastic heterogeneity, such as soft modes[16–18], local yielding stress[19], local thermal energy[20], vibrational mean-squared displacement[21,22], and flexibility volume[23]. These indicators are based on the measurement of physical observables and have clear interpretations, yet typically require detailed knowledge of atomistic interactions. Attempts from a purely structural perspective (*i.e.*, with knowledge of only the atomic positions) have long been frustrated due to the lack of representations to sufficiently encode the structural heterogeneity. Recently, researchers have made notable progress by combining symmetry functions as structural representations with machine learning (ML) to establish predictive models for the plasticity and dynamics of various disordered solids and liquids[24–27].

The use of symmetry functions (originally proposed to fit ML interatomic potentials[28,29]) to establish structure-property relationships in MGs has both advantages and drawbacks. A major advantage is that they can be considered as quite complete, and can successfully distinguish many different types of environments[24–27]. However, the complex and non-intuitive transformations, especially for the angular functions, makes it more challenging to extract scientific insights from ML models employing symmetry functions. Furthermore, to our knowledge no study has demonstrated how ML models employing symmetry functions generalize to different compositions and different chemical systems (*i.e.*, without re-training). As we will later demonstrate, models trained on symmetry functions may be system-specific and limited in their ability to establish more general structure-plasticity mappings that hold across compositions and chemistries.

In this work, we develop a new structural representation by extracting features from the interstice distributions in short- and medium-range that are conceptually related to local susceptibility to rearrangement (Figure 1). We find that this novel representation has advantages over symmetry functions[24–26] as well as conventional signatures[30–38] (e.g., coordination number (CN)[30], Voronoi indices[30], characteristic motifs[31,32], volume metrics[34,35], and $i$-fold symmetry indices[36]). We use these features to explore how the atomic features present in the undeformed, quenched configuration affects plasticity even at large strains and long time scales (see Supplementary Figure 1 for an illustration of differences from previous works). The plastic probability estimates of the ML model, which we call "quench-in softness (QS)", serves as an indicator of the defective nature of site environments and enables us to survey the landscape of soft and hard packings within MGs. Remarkably, we demonstrate that a quench-in softness metric trained on one MG is generalizable across compositions, quenching conditions, and even different chemical systems, suggesting that the traits of atom sites prone to rearrangement could be consistent across different MGs (especially ones containing only metallic elements). Furthermore, the ML framework is general and can be conceivably applied to predict any site-specific property of MGs.

## 2. Results

**Interstice distribution in the short- and medium-range**. To establish a ML link between the site environments and plastic heterogeneity, we must first represent the site environments such that they capture the structural heterogeneity in MGs.

It is well-established that interstices in crystals (also called holes or voids) strongly influence diffusion, deformability and other transport properties[39]. In MGs, however, due to the disordered structure, the interstices are more difficult to define and characterize[40]. In previous studies, Yang et al. have proposed that the atomic packing efficiency, the ratio between the volume of embedded atoms and the total volume of the cluster (equivalent to "1 – interstice fraction"), is strongly correlated with glass-forming-ability of MGs[34]. In terms of plastic deformation, as the rearrangements can be directional and anisotropic, an average interstice fraction alone may be insufficient to distinguish the plastic sites. Indeed, as well will later demonstrate, a more complete representation of interstice distribution, as well as extensions beyond short-range order (SRO) to medium range order (MRO), are required to obtain more accurate models.

In this work, we first characterize the distance, area and tetrahedral volume interstices in the neighboring cluster to construct the SRO descriptors (Figure 1). Each of these metrics is a measure of the relative amount of empty space around each atom as determined using atomic sphere models. The distance interstice is the fraction of a bonding line unoccupied by the atom spheres; it can be negative if the atom spheres overlap. The area interstice is the unoccupied area within the triangulated surface formed by atom triplets in convex hull formed by neighbors, and the volume interstice is the unoccupied portion of a tetrahedra formed between the central atom and neighbor atom triplets; these metrics are typically non-negative. Specifically, we determine neighbors using Voronoi tessellation analysis (an exception is noted later), with small facets with areas smaller than 5% of the average facet areas removed. We then derive the convex hull[41,42], composed by triangulated facets, of the neighbors, to calculate the area and volume interstices. We first calculate the atom-packed area and volume in each triangulated surface and tetrahedron by adding up the circular sector area and cone volumes at each vertex, through calculating the triangular angles and solid angles, and then subtract the atom-packed area and volume from the triangulated facet area and tetrahedron volume to calculate the interstices (Figure 1; see Methods for details).

Iterating the above procedure for all possible interstices in the neighboring environment of each atom will generate 3 vectors whose length is the number of neighbor atoms (distance interstice) or the number of convex hull simplices (area or volume interstice). In essence, the coordination environment of each atom in MGs is anisotropic, which can be reflected in the inequality of the distance, area and volume interstice vector elements. To describe this anisotropy, we derive statistics (mean, min, max and standard deviation) of the interstice vector elements to featurize the interstice distribution around an atom (Figure 1). Other methods can be grouping the interstice vector elements into histogram grids of fixed bins and the features then become a vector of all the values of these histograms (using Gaussian smearing as an option to reduce noise of discrete histogram values and obtain a smoothed distribution). In addition to using Voronoi tessellation to determine the neighbors, the distance interstice metrics can be easily augmented by those calculated from neighbors within a cutoff distance (e.g. 4.0 Å for Cu-Zr MGs), and if so, the number of SRO features will be increased from 12 to 16.

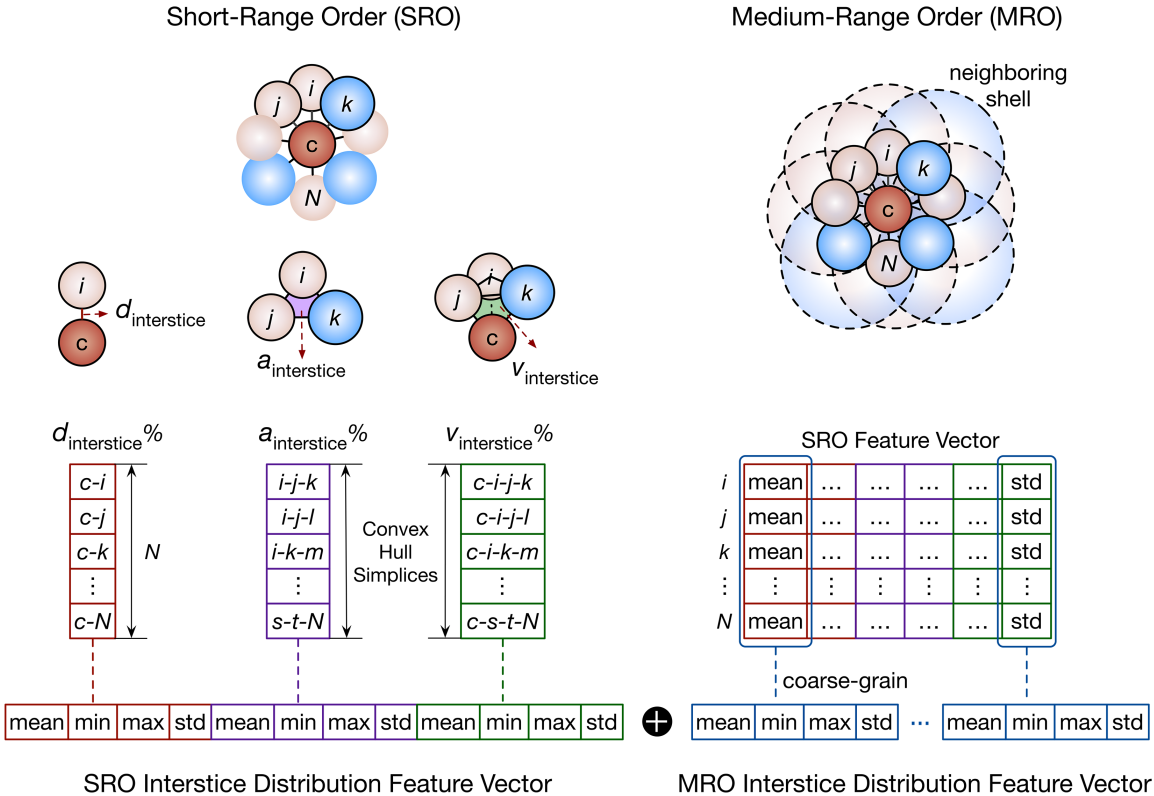

**Figure 1 | Representing site-specific interstice distribution in the short- and medium-range.** We define the distance, area and volume interstice (in ratio) as the fraction of neighboring distances, triangulated surface area and tetrahedra that are not covered by atom spheres. We take statistics (mean, min, max and std) of the distance, area and volume interstices across all nearest-neighbor bonds and convex hull simplices to describe the anisotropy of interstice distribution in the short-range order (SRO) around each atom. The SRO feature vectors of all neighbors can be reduced to a single medium-range order (MRO) feature vector by calculating statistics across neighbors (mean, min, max and std). The SRO and MRO features are then concatenated ($\oplus$) as a representation of the interstice distributions around each atom. The features are then served as input to a ML algorithm (gradient boosting decision tree in this work) to train and predict the heterogeneous plastic response of atoms in MGs.

We next represent the interstice distributions in MRO (Figure 1). Although MRO has long been proposed to be vital to determining glass properties, few MRO signatures are available in literature[43–45]. Here we generalize a coarse-graining strategy to use the statistics of SRO features of an atom's neighbors to describe the center atom itself. Specifically, we process an SRO feature $F^{\text{SRO}}$ to calculate its statistics across the neighbors of atom $i$, and the MRO features will be

$$F_i^{\text{MRO}} = \text{Stats}\bigl(F_1^{\text{SRO}},\ F_2^{\text{SRO}},...,\ F_n^{\text{SRO}}\bigr), \quad n \in N(i) \tag{1}$$

where $n$ iterates the neighbors $N(i)$ determined by Voronoi tessellation analysis or within a cutoff distance, and "Stats" represent the summary statistics of mean, min, max and standard deviation. This allows for automatic encoding of the 2$^{\text{nd}}$ neighbor effect, and is similar to the idea of imposing convolution[46] over neighbors for longer-scale feature extraction. Crucially, this strategy allows us to transform any numeric SRO feature into a set of MRO features. Overall, we see that the SRO and MRO features have clear physical meanings in describing the interstices around each atom, and are robust to varying scales of atomic sizes due to being in the form of ratio.

To summarize, we establish a representation to describe the heterogeneous interstice distribution that spans SRO and MRO around atoms in MGs. Each atom is represented by 80 variables, concatenated from 16 SRO and 64 MRO features ($F = F^{\text{SRO}} \oplus F^{\text{MRO}}$). To reduce model complexity and improve interpretability, we further remove high-linear-correlation features and use recursive feature elimination to reduce the representation to 15 features (listed in Supplementary Table 2, in descending order of the 5-fold cross-validation (CV) averaged feature importances).

The codes for this representation, together with many existing features (such as Voronoi indices, volume metrics, $i$-fold symmetry, bond-orientational order, and symmetry functions), are publicly available in *amlearn* (https://github.com/Qi-max/amlearn), our package targeted for ML in amorphous materials, and *matminer*[47] (https://github.com/hackingmaterials/matminer). In *amlearn*, we wrap Fortran 90 subroutines and functions with Python using f2py[48] to combine usability and fast-computation. This representation is general and can describe the site environments of any MG. We will later show that this representation improves upon the predictive ability of recognized signatures and can even be highly generalizable between different compositions and chemical systems.

**Mapping plastic atoms to quenched-in defects**. Following the "feature extraction" or "fingerprinting" step, we train a ML model to map the features to the property of interest. In our case, this is whether an atom in the quenched structure is susceptible to plastic rearrangement or not. In this work, we select $Cu_xZr_{1-x}$ ($x = 50$, 65 and 80 at.%), which are promising BMG formers[43–45,49–52], as principal alloys and extend analyses to $Ni_{62}Nb_{38}$[53,54], $Al_{90}Sm_{10}$[55] and $Fe_{80}P_{20}$[56] MGs. To generate data for ML, we quench large glass samples (345600 atoms for Cu-Zr and 131072 for other MGs) under quenching rates of $5 \times 10^{10}$, $5 \times 10^{11}$ or $5 \times 10^{12}$ K s$^{-1}$, and apply uniaxial compressive or tensile strain under strain rates of $2.5 \times 10^7$ or $1 \times 10^8$ s$^{-1}$ at 50 K, with periodic boundary conditions along X- and Z- or along all directions, using molecular dynamics simulations (see Methods and Supplementary Figure 2). Here we use gradient-boosted decision tree (GBDT) as the ML algorithm, which builds the prediction model in an iterative manner to construct an ensemble of weak decision tree learners through boosting[57]. To rigorously test the ML models, we quench and compress 3 independent samples for each combination of composition and quenching rate, and use 2 of the 3 samples per condition for training, whereas the 3$^{rd}$ sample is set aside for generalization tests and completely unseen during model development. Due to the imbalanced nature of the datasets, we use equal undersampling for the training data to create a balanced dataset. We then use 5-fold cross-validation (CV) to train a GBDT model that uses the feature vectors of the undeformed configuration to classify atoms that deform plastically up to a strain of 4.0% (see Methods for details). The trained GBDT model is then tested on the completely set-aside generalization sample (thus excluding any trivial information leakage from training) without any undersampling. As a measure of plastic deformation, we use accumulative non-affine displacement[10] ($D^2$) at a relatively large strain (4.0%) with reference to the undeformed configuration, and set a threshold value of 5.0 Å$^2$ to distinguish the plastic and non-plastic atoms. We compare $D^2$ with other plastic indicators in Supplementary Figure 4. The models are compared using their area under receiver operating characteristic curve (AUC-ROC) score (see Methods for the motivation) as well as recall (for consistency with previous works[24–26]).

We begin by discussing $Cu_{65}Zr_{35}$ quenched under a rate of $5 \times 10^{10}$ K s$^{-1}$ as an example (other MGs are discussed later). $Cu_{65}Zr_{35}$ is known as an optimum glass former in the Cu-Zr system[43–45,49–51]. For the ML task of using only the undeformed configuration to predict the plastic atoms accumulated up to a relatively large strain of 4.0%, the AUC-ROC on a set-aside test glass configuration is 0.771, and the ML model captures 74.2% of the true plastic rearrangements (recall). We compare this against baseline models (random, most-frequent and minority predictors) and they give AUC-ROCs of roughly 0.50 (random) or strictly 0.50 (most-frequent and minority), suggesting that such prediction is non-trivial to achieve.

We next test whether our model is improved by adding conventional structural features for disordered solids. We characterize the atoms with another 7 sets of existing geometrical SRO features (CN[30], Voronoi indices[30], characteristic motifs[31,32], volume metrics[34,35], $i$-fold symmetry indices[36] and their weighted version, bond-orientational order[33]) and 1 chemical SRO feature set (numbers of each element type in the neighboring shell, Warren-Cowley parameters[37,38]), totaling 49 SRO features (see Methods), and the AUC does not increase despite the increased number of features (Supplementary Table 5). We also follow the coarse-graining method described above to further generate 10 MRO feature sets, totaling 169 MRO features (Methods), and the AUC has a negligible AUC increase of ~0.002 (Supplementary Table 7). We also test against 166 symmetry functions following the parameters of previous works[24–26] and train a GBDT model with exactly the same data and CV splits. The resulting AUC is 0.751 (Supplementary Table 3), which is slightly lower than but comparable to our result. Furthermore, we will later show that as the formulation of symmetry function is sensitive to length scales, their generalizability to different compositions would be restricted, while our representation and trained ML models exhibit superior generalizability to different compositions and even different chemical systems without re-training the model.

Along with classification, GBDT can determine the probability of each atom to be plastic, which can be considered as an indicator of the plastic susceptibility. For example, a probability of 0.50 indicates the model predicts the atom to have an equal probability to be plastic or non-plastic, and the larger the probability, the greater the likelihood for the atom to be plastic. This is similar to the previously introduced idea of "softness" (distance from the SVM hyperplane)[25,26], but provide (i) well-calibrated probability estimates bounded in the range [0, 1] that can serve as confidence level of classification and does not need further calibration to transform the unbounded SVM distances into probabilities (Supplementary Figure 7) and (ii) determined from the undeformed, quenched configuration immediately after quenching and thus could be considered as "quench-in softness" (denoted as QS).

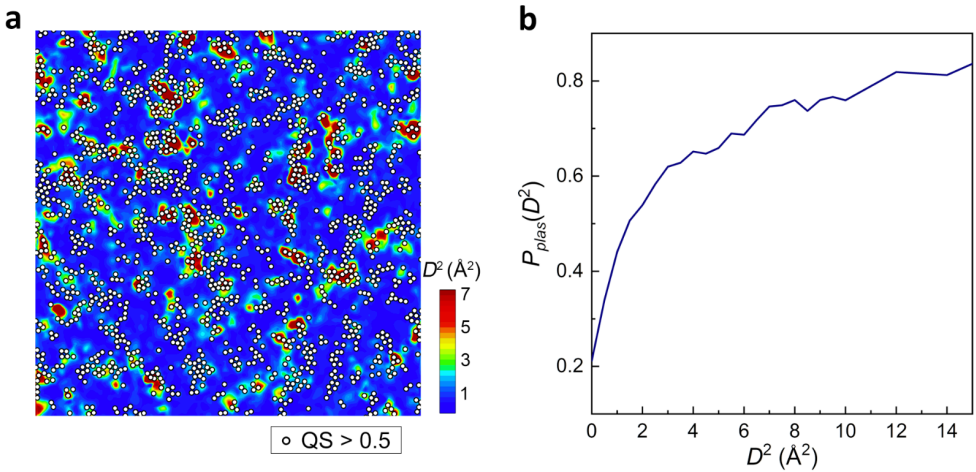

**Figure 2 | Predicting plastic atoms using interstice features and machine learning.** (a) The predicted plastic atoms (white dots) in $Cu_{65}Zr_{35}$ by our ML model versus the distribution of $D^2$ from atomistic simulations in a slice of 2.0 Å thickness at a strain of 4.0% of the prototypical glass of $Cu_{65}Zr_{35}$ quenched at a rate of $5 \times 10^{10}$ K s$^{-1}$. All atoms predicted with "quench-in softness" (QS) larger than 0.5 are marked as white circles. (b) The fraction of atoms predicted as plastic (QS > 0.5) by our ML model (using the undeformed configuration alone as input) as a function of the actual $D^2$ value at a strain of 4.0%.

Figure 2a visualizes the atoms with QS > 0.5, versus the contour map of $D^2$ distribution of the set-aside glass configuration at the strain of 4.0%. Notably, plastic rearrangements have a high propensity to originate from the regions with large QS. The distribution of QS also captures some clustering tendency of plastic atoms, due to the enhanced length scale by incorporating features beyond SRO. Figures 2b shows the likelihood that an atom with an observed value of $D^2$ is predicted to be plastic by ML when given the undeformed configuration as input. This possibility increases with $D^2$, indicating that the more plastic an atom is after applying strain, the more likely it is to be predicted as plastic by the ML model using the initial structure.

The ability to reasonably predict plastic atoms at large strains using the undeformed configuration itself suggests the existence of a long-lived inheritance of plastic heterogeneity on the quenched structure. Here our prediction horizon (strain 4.0%, or equivalently 1.6 ns) from a single structural snapshot is much longer than the previous ML framework (for example, strain 0.02%, or equivalently 400 timesteps[24]). Our model is unique in that, once trained successfully, only a single undeformed snapshot is needed to predict plastic atoms even at relatively large strains. Different from collecting stepwise snapshots to construct the datasets, here our dataset only samples the quenched atomic environment of each atom once (Supplementary Figure 1), and the model is further evaluated with an external test glass configuration that undergoes independent quenching and deformation and thus has different initial configuration and deformation process with the trained configurations.

**Spectrum of QS in metallic glasses**. To understand the variety of site environments present within a MG, we examine the distribution of quench in softness (Figure 3a). A long-tail in the higher QS (soft) end is observed, while the low QS (hard) end distributes more smoothly. We further plot the probability that an atom rearranges as a function of QS[26] (Figure 3b). This probability is a strong function of QS, increasing by several orders of magnitude from the hardest to the softest atoms. Furthermore, a value of QS = 0.5 corresponds to a plastic likelihood that is equal to the overall fraction of plastic atoms. This is demonstrated in the right-side axis of Figure 3b, where the quantity P(plastic|QS)/P(plastic) is close to 1.0 for QS = 0.5. In the lower QS (hard) end, the curve bends at ~0.1, below which the atoms are at least ~10 times less probable to be plastic than average. These atoms cover ~13% of the total atoms and could be viewed as the "hardest" or most "solid-like" atoms. Their average $D^2$ is ~0.55, suggesting they mostly respond elastically with minor non-affine rearrangement. In the soft end, we consider atoms with QS > 0.7 (the beginning of the soft "tail") as the "softest", or most "liquid-like" atoms, with a similar atomic fraction of ~11%.

We proceed to address how these characteristic atoms pack in space. We first perform fractal dimensionality sampling[58,59] for the hardest and softest atoms using the power-law scaling of the mass distribution $M(r) \sim r^D$, where $M(r)$ denotes the number of atoms of each type within radius $r$ centered by an atom (Figure 3c). Theoretically, the slope $D$ of the $M(r)$ curve in log-log plot is the dimensionality, and $D < 3$ indicates fractal structure[58,59] (*i.e.*, the number of atoms does not straightforwardly increase with the volume of an enclosing sphere). We see the hardest and soft atoms both show fractal-like packing in length scales below 10 Å (~ 4 neighboring shells), beyond which the packing becomes more space-filling ($D$ close to 3). The fractal-like characteristics is much stronger in the hardest atoms than the softest ones. We further extract the pair correlation functions $g(r)$ of the two groups of atoms (Figure 3c inset). The 1$^{st}$ peak of the hardest atoms is higher, suggesting a higher neighboring tendency. Beyond the neighboring shell, the hardest atoms still exhibit clear coordination peaks up to ~ 4 neighboring shells, while the peaks of softest atoms quickly smear out. The distinct coordination behaviors suggest that the hardest atoms are more likely to form a plastic-resistant backbone that penetrate in MGs, while the soft spots are essentially localized in space without extending beyond SRO.

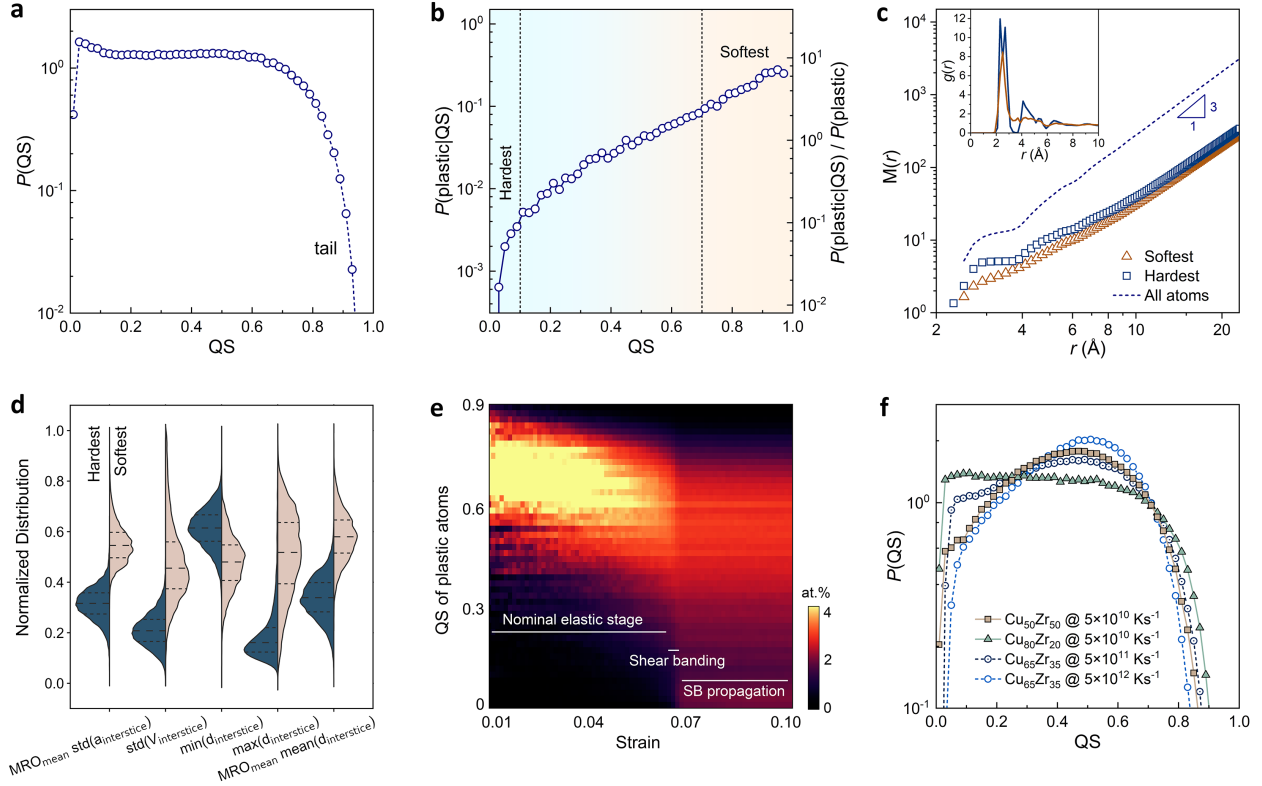

**Figure 3 | Distribution of machine-learnt quench-in softness in metallic glasses.** (a) Wide spread in the distribution of quench-in softness (QS) in $Cu_{65}Zr_{35}$ defines a distribution of atomic environments within a glass. (b) The probability that an atom rearranges as a function of QS, $P(plastic|QS)$, is a strong function of QS, increasing by several orders of magnitude, from hardest atoms at low QS end to softest atoms with high QS. (c) Fractal dimensionality sampling in a log-log plot for the hardest (QS < 0.1) and softest atoms (QS > 0.7) as well as all atoms, based on the power-law scaling of the mass distribution $M(r) \sim r^D$, where $M(r)$ denotes the number of atoms within radius $r$ centered by an atom and slope $D$ is the dimensionality. The inset shows the pair correlation function functions $g(r)$ of the two groups of atoms. (d) Violin plots of representative interstice distributions of the hardest and softest atoms, manifesting strong site environment contrasts between these two groups of characteristic atoms. (e) QS of the plastic atoms activated under a certain strain. The entire deformation range up to the strain of 10% with an interval of 0.125% is analyzed. The crossover between nominal elastic stage (plasticity highly dependent on high-QS atoms), and shear banding and shear band propagation stages (plastic rearrangement extending across all levels of QS) is clearly shown and marked. (f) QS distributions of $Cu_{50}Zr_{50}$ @ $5 \times 10^{10}$ K s$^{-1}$, $Cu_{80}Zr_{20}$ @ $5 \times 10^{10}$ K s$^{-1}$, $Cu_{65}Zr_{35}$ @ $5 \times 10^{11}$ K s$^{-1}$ and $Cu_{65}Zr_{35}$ @ $5 \times 10^{12}$ K s$^{-1}$ to reveal the effects of composition or thermal history. By varying these parameters, one can in principle tailor the site-specific plastic response of a MG.

Next, we examine the structural traits of the hardest and softest atoms (Figure 3d). Strong contrasts in the site environments are observed for these two groups of atoms, and a large degree of separation can be achieved even with a single interstice distribution feature. For example, the softest atoms typically have less regular, more anisotropic neighboring environments with high variance in the distance, area and volume interstices, and this effect is more pronounced when such anisotropy is present at both short- and medium-ranges. As another signature, atoms with extremely low minimum bond interstice (this typically means a bond distance smaller than equilibrium distance, *i.e.*, in the repulsive regime) and large maximum bond interstice (atoms too far apart) in the neighboring shell are more prone to be soft. In previous studies, there are two major approaches to establish the structure-plasticity relations in MGs: one focuses on the identification of locally-favored atoms that are resistant to plastic deformation and behave as elastic backbone of MGs, and the other focuses on identifying flow-defects, or soft spots that are plastic carriers. The machine-learnt QS encompasses both ends of the spectrum and provides a complete landscape of structural deformability, from the hardest end to the softest end, in MGs.

In practice, atoms frozen in the quenched structure are gradually activated plastically with the increase of strain. We trace the QS of the activated plastic atoms as the strain is applied (Figure 3e). We see the progression of plastic sites indeed follows a sequence. At low strains, the plastic atoms correspond mainly to those predicted with high QS, *i.e.*, high probability predicted by ML to be deformable. As the strain progresses, plasticity is induced at sites predicted with gradually lower QS. Thus, less susceptible sites are essentially frozen until the stress is large enough to trigger the rearrangement. Throughout, a fraction of low QS atoms are also activated due to inevitable stochastic effects and shear avalanches.[14] Yet, during this entire range, the QS reasonably distinguishes plastic and non-plastic atoms. However, we also see that this sequence is abruptly disrupted by shear banding at a strain of ~0.065. Significant plastic rearrangement avalanches occur near yielding (a local rearrangement triggers others, leading to a cascade)[12], and the atoms that form shear bands cannot be identified as pre-existing structural defects. Upon shear banding, plastic rearrangement abruptly extends across all levels of quench-in softness QS (Figure 3e), suggesting that a largely environment-independent transformation occurs along the pathway of shear bands (typical shear banding snapshots are shown in Supplementary Figure 3).

Thus, deformation occurs first in regions of high QS followed by those with low QS, and we note that QS appears to be a relevant metric only prior to the formation of shear bands.

Finally, we characterize how the QS distribution is affected by the glass composition its thermal history (Figure 3f). $Cu_{80}Zr_{20}$ has a similar QS distribution to that of $Cu_{65}Zr_{35}$ (Figure 3a), while $Cu_{50}Zr_{50}$ is more centered around QS of 0.5, suggesting the site environments are less heterogeneous. The fraction of hardest atoms in $Cu_{50}Zr_{50}$ is also notably lower (~5.4% atoms with QS < 0.1), suggesting a lower fraction of exceptionally plastic resistant atoms within the glass. We then calculate the standard deviation of QS, *i.e.* std(QS), as an indication of structural heterogeneity, and the std(QS) of $Cu_{50}Zr_{50}$, $Cu_{65}Zr_{35}$ and $Cu_{80}Zr_{20}$ is 0.196, 0.230 and 0.234, respectively. This agrees with previous studies suggesting $Cu_{50}Zr_{50}$ does have a lower structural and plastic heterogeneity than $Cu_{65}Zr_{35}$[43,44]. As to thermal history, with the increasing quenching rate, QS variation also gradually decreases (Figure 3f), indicating a lowered structural heterogeneity developed during quenching. The fraction of hard atoms also decreases. Overall, faster quenching results in lower structural heterogeneity that should be closer to the parent liquids.

**Generalization to new compositions, quench rates and systems**. Thus far, we have trained our ML model for a specific glass system and tested it on unseen glass configurations under the same condition, *i.e.*, same composition and quench rate. This is already a more robust generalization test beyond the traditional train-to-test within a single glass configuration. Extrapolating even further, there is a more challenging yet significant question - is it possible for the ML models to generalize across different compositions, thermal histories and even different chemical systems, without re-training? This type of test has not been performed in previous glass studies[24–26] and is generally much more challenging for all categories of ML studies[60].

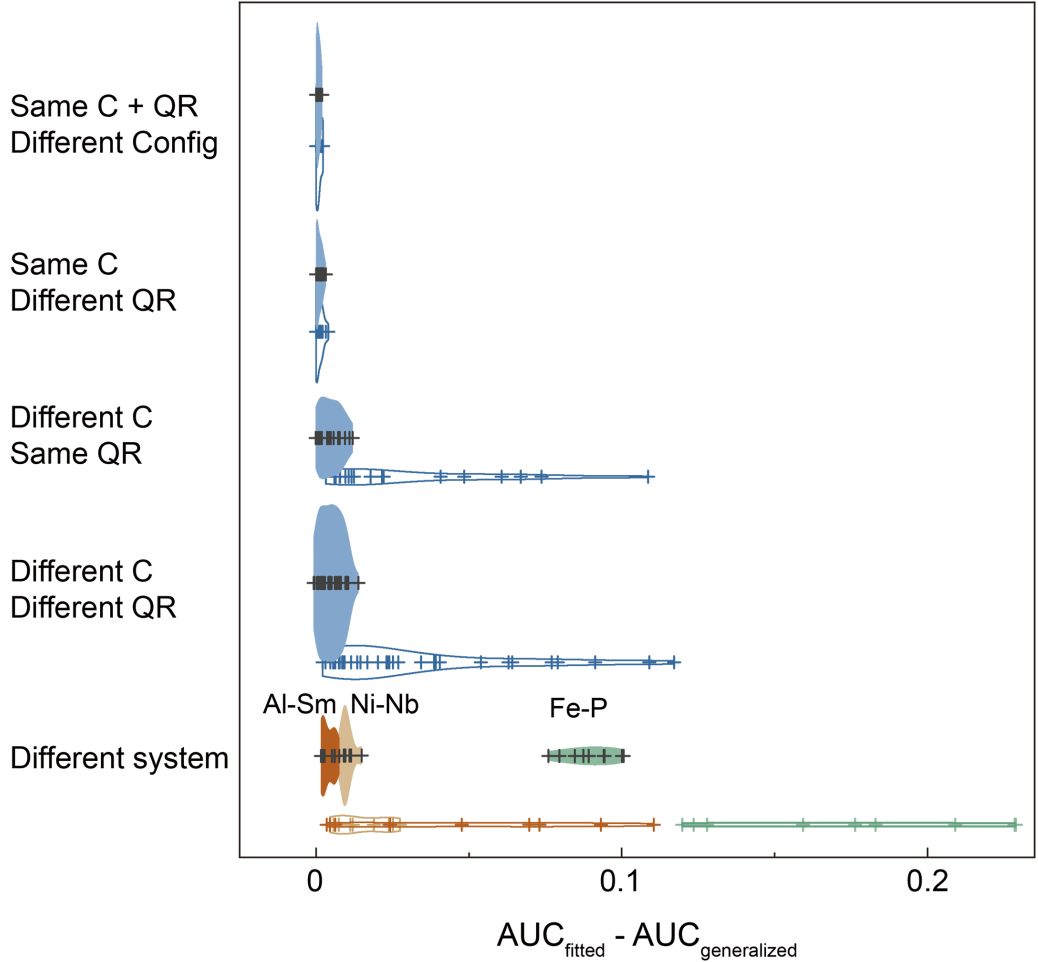

**Figure 4 | Generalizability of machine learning models**. Generalization performances of the ML models using our interstice distribution features (filled violin plots) and those using symmetry functions (unfilled violin plots) on 5 generalization scenarios: i) same composition (abbreviated as C) and same quenching rate (QR) but to new unseen configurations, ii) same composition and different quenching rate; iii) different composition and same quenching rate; iv) different composition and different quenching rate; v) different system. The performance is evaluated by the difference of the generalization score and the original, same-condition score. The height of the violin plot is proportional to number of tests that fall within a performance interval and the interior shows the data points.

As a first test of generalization ability, we stayed within the same chemical system and tested all 81 possible mutual generalization pairs between the 9 Cu-Zr MGs with varying compositions and quenching rates. The tests can be generally grouped into 4 categories: i) generalization to unseen glass configurations with same composition and quenching rate (9 tests), as a reference; ii) generalization between glasses with the same composition but different quenching rate (18 tests); iii) same quenching rate but different composition (18 tests); iv) different composition (same chemical system) and quenching rate (36 tests). The generalization performance is evaluated by the difference between the AUC of a model trained specifically on a target glass with the generalized AUC achieved by applying a ML model trained for another glass to the target glass (filled violin plots in Figure 4). Interestingly, for the case of our ML framework, we see that the generalization performances are quite close to the fitted cases for all $1^{st}$ to $4^{th}$ scenarios (the $3^{rd}$ and $4^{th}$, *i.e.* transferring between different Cu-Zr compositions, are slightly worse), with AUC decreases of <0.015 (see Supplementary Table 9 for typical fitted and generalization scores). This suggests a strong generalizability of our feature representation and obtained ML models between MGs within a single system.

As a more difficult problem, we test the generalizability of our learnt models to completely different chemical systems. In addition to Cu-Zr MGs, we extend our ML studies to $Ni_{62}Nb_{38}$, $Al_{90}Sm_{10}$ and a metal-metalloid glass $Fe_{80}P_{20}$ (Methods). When directly training models on these systems, we achieve AUCs of 0.737 - 0.775 for predicting the plastic atoms during tensile or compressive deformation (see Supplementary Table 8). The comparable accuracy suggests our interstice representation and ML framework can apply to MGs of various structural traits. We next tested the 27 generalization pairs from the 9 Cu-Zr MGs to $Ni_{62}Nb_{38}$, $Al_{90}Sm_{10}$ and $Fe_{80}P_{20}$ MGs. These tests form the $5^{th}$ generalization category: v) different chemical system (27 tests). As seen from Figure 4, even when generalizing between different chemical systems, the models fitted in Cu-Zr MGs can even achieve good performances in the Ni-Nb and Al-Sm MGs with minor loss of AUCs, suggesting that the relative rankings of QS by models learnt in different MGs can be similar. The generalization to Fe-P MG is worse (with AUC decreases of ~0.07 - 0.10), which is expected as the decisive features in distinguishing the plastic and non-plastic atoms in the metal-metalloid MG are likely to be different from that of the all-metal MGs (Supplementary Table 10).

As a direct comparison, we have calculated the symmetry functions for our studied MGs based on the formulation and parameter settings of the previous works[24–26] (Methods), and train GBDT models on the same data and CV splits with optimized hyperparameters. As discussed above for the Cu-Zr MGs, the symmetry functions overall achieved comparable AUCs (mostly with ~0.01 - 0.02 lower) with our features (Supplementary Tables 3 and 8) when directly training models on a specific glass. Nonetheless, when transferring the ML models, the generalizability of the two sets of models is quite different. For the ML models using symmetry functions as input, the loss in AUC on the $1^{st}$ and $2^{nd}$ generalization tests is minor (also see Supplementary Table 9 for typical fitted and generalization scores) and comparable to our method (unfilled violin plots in Figure 4). However, when attempting to generalize to different compositions (*i.e.*, $3^{rd}$ and $4^{th}$ tests), the AUC degrades substantially. The degradation is even worse when generalizing between different chemical systems (except for generalizing from Cu-Zr to Ni-Nb). Thus, ML models trained using symmetry functions have a limited ability to describe other systems. We hypothesize that this is because symmetry functions (and therefore the trained ML models) are by definition more sensitive to the length scales and coordination environments of the specific system. The radial part of the symmetry functions (often more important than the angular ones) characterizes the Gaussian-smeared radial density of each species at a series of distances (Methods). Although the distances are often normalized to the equilibrium distance between one species[24–26], the radial functions can still be very different among systems with distinct atomic sizes and coordination environments.

**Revisiting previously proposed signatures and rules**. As discussed above, despite many recognized structural signatures have been proposed for MGs[30–38], a standard way is still lacking to quantitatively assess and compare the predictive ability of the signatures as well as the structure-property correlation proposed.

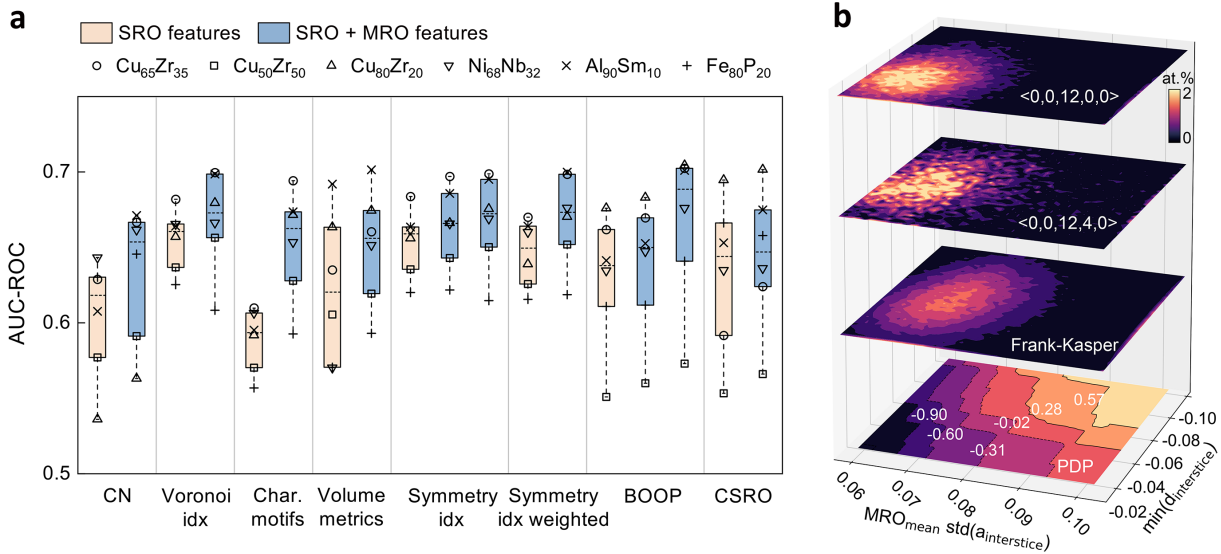

**Figure 5 | Quantitatively benchmarking individual feature sets**. (a) Quantitatively benchmarked predictive powers of individual short- (SRO) feature sets and the feature sets further augmented by the medium-range order (MRO) features generated following the coarse-graining technique described in this work. Please refer to Methods and Supplementary Tables 4 and 6 for a full description of the feature sets. Results for $Cu_{65}Zr_{35}$, $Cu_{50}Zr_{50}$, $Cu_{80}Zr_{20}$, $Ni_{62}Nb_{38}$, $Al_{90}Sm_{10}$ and $Fe_{80}P_{20}$ MGs (quenched under $5 \times 10^{10}$ K s$^{-1}$) are shown. In the box plots, bounds of the box spans from 25 to 75% percentile, dashed line represents median, and whiskers show minima and maxima of data points. (b) Projecting <0,0,12,0,0>, <0,0,12,4,0> and Frank-Kasper clusters to the two-dimensional (2-D) partial dependence plot (PDP) of the top-2 features of the ML model. Conventional descriptors only capture a small fraction of the possible input space whereas our features and ML form a more complete description of feature space.

In this work, we demonstrate that ML can be used as a tool to quantitatively assess the predictive capability of candidate feature sets.[61] By feeding each individual feature set to train a ML model, the prediction score on the same datasets and CV splits can be used as a metric of the feature set's predictive ability (Figure 5a). Here we consider 8 existing SRO feature sets and those sets further augmented by the MRO features generated following the coarse-graining scheme described above (Please refer to Methods and Supplementary Tables 4 and 6 for a full description of the feature sets). Results indicate that individual feature sets yield prediction accuracies varying notably among MGs, even among different compositions within a single Cu-Zr system. Overall, the plastic atoms in $Cu_{50}Zr_{50}$ and $Fe_{80}P_{20}$ are the most difficult to predict for these feature sets. Augmenting the SRO features by the coarse-grained MRO features overall improves the predictions.

Interestingly, we find that the SRO Characteristic motifs set that describe whether an atom's neighbors form icosahedra <0,0,12,0,0> (or <0,0,12,0> if omitting facets with 7 edges and more), <0,0,12,4,0>, or Frank-Kasper clusters[31] generally have the lowest AUCs among all SRO feature sets. Despite these clusters have been identified to be of the most stable clusters in MGs, especially the Cu-Zr MGs, not forming any of these clusters does not mean that the atom is not stable. Meanwhile, as clusters with the same Voronoi indices could have different packing, falling in one of these clusters does not guarantee a low plastic susceptibility. Both factors restrict the plastic/non-plastic distinguishability of these motifs. However, when extended to MRO, the predictive power of the characteristic motifs features has greatly enhanced, e.g. AUC increasing from 0.610 (SRO) to 0.694 (SRO+MRO) for $Cu_{65}Zr_{35}$. This evidences that the cluster-cluster connection of these motifs in the medium-range could be more important, as discussed in previous studies[43,44].

We further plot the 2-dimensional partial dependence plot (PDP)[57] of our top-2 features, and overlay the positions in that feature space for atom sites corresponding to <0,0,12,0,0> icosahedra, <0,0,12,4,0> and Frank-Kasper clusters (Figure 5b). Due to the difficulty of visualizing the decision boundary in high-dimensional feature space, PDP offers a mechanism to resolve the effect of specific features by marginalizing the model output over other features, in essence measuring how the model prediction changes (on average) as a function of the target features[57]. We see that these characteristic motifs are indeed residing in the regions that correspond to low plastic susceptibility (negative partial dependence). This confirms the increased stability of these long-proposed motifs. Nonetheless, all the motifs are distributed only in a narrow portion of the entire feature space, while in contrast, our features form a more complete description of feature space (Figure 5b). These interstice features (together with the other 13 features not shown) capture not only the stable nature of regions of feature space corresponding to these characteristic motifs, but also the effect of less conventional feature space (e.g. unstable SRO but stable MRO may also stabilize the atoms) to form the final decision boundary. We refer to Supplementary Figures 8 - 25 for a more complete analyses of PDPs for interpreting each SRO or MRO feature set. Furthermore, we reiterate that further augmenting our interstice representation with these SRO and generated MRO feature sets leads to a negligible AUC increase (Supplementary Tables 5 and 7).

## 3. Discussion

The present work can also have impact on the design of MGs with tailored mechanical behaviors. Our findings demonstrate that much of plasticity within the elastic regime (prior to shear band formation) is controlled by QS that is determined by the initial glass configuration, rather than by complex dynamics. Thus, any processing route that can modify the glass structure could potentially tailor the distribution of QS in the materials (e.g., as in Figure 3f) and thus tailor the resultant deformation responses. For instance, it has been proposed that to make MGs ductile, a larger density of plastic units is preferred, and plausible routes can be changing the thermal history, such as ultrafast quenching, rejuvenating glass structure through thermomechanical cycling, or applying irradiation techniques to transform the glass into more a deformable structural state[62]. As our ML models can generalize well to unseen glass configurations, even of different chemical systems, one can use the models to get some quick estimates on the plastic heterogeneity of candidate configurations derived by processing routes of interest.

Furthermore, the generalizability of our models suggest they can be applied even in the absence of simulated configurations tailored to the specific system of interest, and could be applied directly to glass configurations inversely generated from experimental diffraction or EXAFS spectra using techniques such as reverse Monte Carlo (RMC)[34]. Herein, the use of ML could help accelerate the development cycle of glasses with targeted mechanical behaviors. We also note that the ML-learnt mapping between the interstice distribution and plastic heterogeneity are much more generalizable between MGs that only contain metallic elements than to metallic-metalloid MGs (Figure 4). This can be attributed to their different atomic interactions: one principally includes metallic bonds and another involve directional covalent bond contributions. Thus, generalization performance should be expected to vary depending on the type of system under investigation.

In this work, we use data from a single compressive/tensile deformation simulation to fit the ML models. We note that using athermal quasi-static deformation[63] or iso-configurational ensemble technique[21] to construct the datasets may reduce thermal fluctuations and stochastic effects and improve the prediction scores. We choose to use a single deformation simulation to mimic the real experimental deformation in which thermal fluctuations and stochastic effects do play a role. Here, we show that ML is capable of learning a mapping that best explains the data (even if there are some noises inside), and the learnt model can even be generalized to completely unseen compositions or chemical systems with appropriate representation and learning protocols.

To summarize, the heterogeneity of atomic environments in MGs makes it formidably challenging to predict their response to external stimuli at the atomic-scale. In this work, we demonstrate that focusing on the short- and medium-range distribution of interstitial spaces (distances, areas and volumes) and applying ML can help form an interpretable and generalizable model to predict the atomic-scale response to mechanical stress for several different systems. In addition to deformation, the ML framework we describe is readily generalizable to the studies of other site-dependent properties and could also be applied to other important physical processes such as thermal activation, glass transition and relaxation.

## Methods

**Featurizing Interstice Distribution in Metallic Glasses**. We use two methods for determining near-neighbors: Voronoi tessellation and cutoff distances. The convex hull is derived using scipy[41] (based on qhull library[42]) with qhull option of "Qt" (triangulated output), and all facets will be simplicial.

The procedure of calculating the distance interstice between center atom O and neighbor A: i) calculate the distance $d_{bond}$ between O and A; ii) calculate atom-packed distance $d_{pack}$ as the sum of atom sizes as $\sum_{O,A} R_i$, where $R_i$ is radius of the atom at site $i$; iii) derive the distance interstice as $(d_{bond} - d_{pack})/d_{bond}$.

The procedure of deriving the area interstice of facet ABC: i) calculate the triangle area $a_{triangle}$; ii) calculate the angle $\theta_i$ of each vertex $i$ as $\arccos(\frac{\mathbf{r}_{ij} \cdot \mathbf{r}_{ik}}{|\mathbf{r}_{ij}||\mathbf{r}_{ik}|})$; iii) calculate the atom-packed circular sector area $a_{pack}$ as $\sum_{A,B,C} R_i^2 \theta_i/2$; iv) derive the area interstice as $(a_{triangle} - a_{pack})/a_{triangle}$.

The procedure of computing the volume interstice of tetrahedron formed by center atom O with facet ABC: i) calculate the tetrahedron volume $v_{tetrahedron}$; ii) calculate the solid angle $\Omega_i$ of each vertex $i$ as $2\arctan(\frac{\mathbf{r}_{ij} \cdot (\mathbf{r}_{ik} \times \mathbf{r}_{il})}{|\mathbf{r}_{ij}||\mathbf{r}_{ik}||\mathbf{r}_{il}| + (\mathbf{r}_{ij} \cdot \mathbf{r}_{ik})|\mathbf{r}_{il}| + (\mathbf{r}_{ij} \cdot \mathbf{r}_{il})|\mathbf{r}_{ik}| + (\mathbf{r}_{ik} \cdot \mathbf{r}_{il})|\mathbf{r}_{ij}|})$, with care of arctan to avoid negative value; iii) calculate the atom-packed cone volume $v_{pack}$ as $\sum_{O,A,B,C} R_i^3 \Omega_i/3$; iv) derive the volume interstice as $(v_{tetrahedron} - v_{pack})/v_{tetrahedron}$.

In this work, we use atomic radii from Miracle et al[35]. One can also use the equilibrium distance estimates from pair correlation functions or other sources. The values of interstices would be affected by the atomic radii, but this will not affect the performance of ML as long as the classes are distinguishable.

In essence, this representation is also applicable to crystalline interstices. As an example, for a one-component bcc structure, supposing that each atom perfectly touch the 8 neighbors, the distance, area and volume interstice vector would be $[0, ..., 0]_{length=8}$, $[0.41, ..., 0.41]_{length=12}$, and $[0.32, ..., 0.32]_{length=12}$. It follows that the mean, min and max of the distance, area and volume interstice distribution features will be 0, 0.41 and 0.32 (it is known that the volumetric packing factor of a perfect bcc structure is 0.68), respectively, and the standard deviations will all be 0.

**Machine Learning**. We use gradient boosting decision trees (GBDT) as our ML algorithm and the hyperparameters searched in this work can be found in Supplementary Table 1. The GBDT model is trained on the plastic heterogeneity data at a strain of 4.0% to learn to classify the atomic environments back in the undeformed configuration as "plastic" or "non-plastic". For Cu-Zr MGs, a single model is fitted for both Cu and Zr atoms, and for $Ni_{62}Nb_{38}$, $Al_{90}Sm_{10}$ and $Fe_{80}P_{20}$, the model is for the host (majority) atoms only. Due to the localized plasticity of glasses under low temperatures or slow strain rates, the non-plastic atoms heavily outnumber the plastic atoms (approximately 3.5% – 6.0% of atoms are plastic at a strain of 4.0%). We deal with the between-class imbalance by random equal undersampling to create a balanced dataset. After performing 5-fold cross-validation on the sampled datasets, we generalize the obtained models to the unseen glass configuration and calculate the average scoring metric. In this work, we use area under receiver operating characteristic curve (AUC-ROC) on the unseen glass configuration as the scoring metric, instead of the recall used in previous studies (although we also report recall for comparison purposes when needed).

It is worth noting that we propose to report AUC along with the recall used in previous works because i) AUC evaluates the tradeoff between true positive rate against false positive rate as a function of chosen threshold, and thus balances the tradeoff between over- and under-predicting the plastic atoms, and does not depend on optimizing a specific prediction threshold (as with precision, recall, or f1 score); ii) the AUC score can be interpreted as the probability that a true positive atom (plastic) is assigned a higher plastic probability (ranks higher than) a true negative atom (non-plastic).[64] This is in accord with the scenario suggested by glass dynamics, which indicates that many structurally soft atoms in the glasses may not be activated under each deformation test, and thus a good model would aim to increase the possibility that the activated plastic atoms are ranked higher than elastic ones. iii) AUC-ROC is robust with the imbalanced data[65] (see Supplementary Figure 6 for an illustration). We note that we prefer AUC-ROC over area under the precision-recall curve because AUC-ROC gives equal weight to plastic and non-plastic atom classification.

**Symmetry Functions**. Symmetry functions are first proposed to fit ML interatomic potential[28,29], and later employed to represent the atomic environment in disordered materials[24–26]. For an atom $i$ in a binary A-B system, the radial and angular symmetry functions are described as[24–26],

$$G_X(i\,;\,r) = \sum_{j \in X} e^{-(r_{ij} - r)^2/2\sigma^2} \tag{2}$$

$$\psi_{XY}(i\,;\xi, \lambda, \zeta) = \sum_{j \in X}\sum_{k \in Y} e^{-(r_{ij}^2 + r_{ik}^2 + r_{jk}^2)/\xi^2}(1 + \lambda\cos\theta_{ijk})^\zeta \tag{3}$$

here $X$, $Y$ denote the atom species in the system: in Eq. 2, $X$ can be A or B, and in Eq. 3,

($X$, $Y$) can be (A, A), (A, B) or (B, B). $r_{ij}$ is the distance between atoms $i$ and $j$, $\theta_{ijk}$ is the angle between $\mathbf{r}_{ij}$ and $\mathbf{r}_{ik}$, $\sigma$ is a constant that is often set as the bin size of $r$, and $r$, $\xi$, $\lambda$, and $\zeta$ are variable constants. The sums are taken over atom pairs whose distance is within a large cutoff $R^c$.

The radial part characterizes the Gaussian-smeared radial density of species $X$ at each $r$ (proportional to $r^2 g(r)$ if $\sigma \to 0$, where $g(r)$ is the radial density function around atom $i$), and the angular part characterizes bond orientations following the formulation of Behler et al[28]. Varying $r$, $\xi$, $\lambda$, and $\zeta$ generates a group of symmetry functions that characterize the site environments around each atom[24–26]. In this work, we follow the settings of $r$, $\xi$, $\lambda$, and $\zeta$ in previous works[24–26] to derive the symmetry. For example, for an atom $i$ in Cu-Zr MGs, we derive 100 radial functions (50 for $i$-Cu and 50 for $i$-Zr) by varying $r$ from 0 to 5.0 × Cu-Cu equilibrium distance (sum of metallic radii) with increments of 0.1 × Cu-Cu equilibrium distance, and 66 angular functions (22 for Cu-$i$-Cu, Cu-$i$-Zr and Zr-$i$-Zr, respectively) by using 22 sets of $\xi$, $\lambda$, and $\zeta$ for each atom. Please refer to the original papers[24–26] for details of parameter settings.

**Benchmarked Structural Signatures and Their MRO Version.** To conduct an extensive benchmarking of candidate feature sets in the field of MGs (Figure 5a), we have featurized 8 SRO feature sets, and further extract their MRO features following the coarse-graining technique described above (Eq. 1), and combine them with the SRO features to explore their predictive capability if being extended from SRO to MRO. The SRO feature sets as well as the statistical types used in generating the MRO ones are listed as follows. Some feature sets such as $i$-fold symmetry and BOOP have two ways to extend to MRO, and both are included in the benchmarks. The order is the same with that (from left to right) in Figure 5a, with the number of features in parentheses.

i)  *CN* (2): Coordination number by Voronoi tessellation[30] or by cutoff distance; *MRO CN$_{Voro/Dist}$* (8): mean, std, min and max;

ii)  *Voronoi idx$_{3…7}$* (5): $\{n_i\}$ where $n_i$ is the number of $i$-edged facets ($i$ in the range of 3-7) in the Voronoi polyhedra[30];

*MRO Voronoi idx$_{3...7}$* (20): mean, std, min and max;

iii) *Characteristic motifs* (4): One-hot encoded signatures of whether a cluster belongs to <0,0,12,0,0>, <0,0,12,4,0>, <0,0,12,0,0>||<0,0,12,4,0> or Frank-Kasper-type clusters[31];

*MRO Characteristic motifs* (12): sum, mean and std (min and max are not helpful, as they are one-hot encoded features and the min and max over the neighbors would be 0 and 1 in almost all cases);

iv) *Volume metrics* (3): Cluster packing efficiency[34], atomic packing efficiency[35] and the ratio of the atomic volume to the Voronoi polyhedron volume around each site;

*MRO Volume metrics* (12): mean, std, min and max;

v) *i-fold symm idx$_{3...7}$* (5): $n_i/\sum_{i=3}^{7} n_i$ where $n_i$ is Voronoi index ($i$ in the range of 3-7), reflecting the strength of $i$-fold symmetry in local sites[36];

*MRO Avg. i-fold symm idx$_{3...7}$* (5): $\sum_{m=0}^{\mathrm{NN}} n_i^m / \sum_{m=0}^{\mathrm{NN}} \sum_{i=3}^{7} n_i^m$, where $n_i$ denotes the number of $i$-edged facets of the Voronoi polyhedra and $m$ iterates over each neighbor;

*MRO i-fold symm idx$_{3...7}$* (20): mean, std, min and max;

vi) *weighted i-fold symm idx$_{3...7}$* (5): using Voronoi facet areas as weights in calculating the $i$-fold symm idx$_{3...7}$;

*MRO Weighted i-fold symm idx$_{3...7}$* (20): mean, std, min and max;

vii) *BOOP $q_{4...10\text{-Voro/Dist}}$ and $w_{4...10\text{-Voro/Dist}}$* (16): Lowest- and higher-order rotation-invariant $q_l$ and $w_l$ ($l = 4, 6, 8$ and $10$) of the $l$th moment in a multipole expansion of the bond vector distribution on a unit sphere[33];

*Coarse-grained BOOP* (16): Coarse-grained[66] lowest-order and higher-order rotation-invariant $\bar{q_l}$ and $\bar{w_l}$ ($l = 4, 6, 8$ and $10$).

*MRO BOOP $q_{4...10\text{-Voro/Dist}}$ and $w_{4...10\text{-Voro/Dist}}$* (64): mean, std, min and max;

viii) *CSRO$_{Voro/Dist}$* (9): Element type, the number and the deviation of local chemistry with nominal composition (Warren-Cowley parameters[37,38]).

*CMRO$_{Voro/Dist}$* (32): mean, std, min and max.

These SRO features are among the most recognized signatures in the field of MGs. Voronoi tessellation (signified by subscript "Voro") and cutoff distance (subscript "Dist") are both used to define neighbors in calculating CN, BOOP, distance statistics and CSRO.

One can also refer to Supplementary Tables 4 and 6 for a more detailed description of the features. These SRO features can be calculated from *amlearn* and the MRO features can be derived using helper statistical functions in *amlearn* or *matminer*.

**Liquid Quenching and Deformation Simulation**. We simulate liquid melt quenching and deformation of $Cu_xZr_{1-x}$ ($x$ = 50, 65 and 80 at.%), $Ni_{62}Nb_{38}$, $Al_{90}Sm_{10}$ and $Fe_{80}P_{20}$ MGs using molecular dynamics simulations. We use 3 quenching rates of $5 \times 10^{10}$, $5 \times 10^{11}$ and $5 \times 10^{12}$ K s$^{-1}$ for Cu-Zr MGs, and $5 \times 10^{10}$ K s$^{-1}$ for all other MGs. We construct 3 large slab samples for each Cu-Zr MG, each of which contains 345600 atoms with dimensions ~120 ($X$) × 24 ($Y$) ×240 ($Z$) Å$^3$. Data from 2 glass samples are concatenated, equally undersampled and used in 5-fold cross-validation training the ML models, whereas the remaining sample is set-aside for rigorous generalization tests. For $Ni_{62}Nb_{38}$, $Al_{90}Sm_{10}$ and $Fe_{80}P_{20}$ MGs, we construct samples of 131072 atoms. We use LAMMPS[67] and EAM potentials in Ref.[52,54–56] The timestep is 1 fs. During simulation, the initial configuration is built by randomly substituting into an fcc (Cu-Zr, $Ni_{62}Nb_{38}$ and $Al_{90}Sm_{10}$) or bcc ($Fe_{80}P_{20}$) lattice. The samples are annealed at 2000 K for 1 ns, quenched to 50 K with each quenching rate, and relaxed at 50 K for 1 ns.

After quenching, the Cu-Zr MGs are compressed along $Z$ axis under a strain rate of $2.5 \times 10^7$ s$^{-1}$ in a quasi-static mode (constantly apply a small strain and then relax, up to the strain of 10%) at a low temperature of 50 K (see Supplementary Figure 3 for typical stress-strain curves). Periodic boundary conditions (PBCs) are imposed in $Y$ and $Z$ axes and free surfaces are applied along $X$ axis to allow shear offsets. For $Ni_{62}Nb_{38}$, $Al_{90}Sm_{10}$ and $Fe_{80}P_{20}$, we simulate both tensile and compressive deformation with strain rates of $2.5 \times 10^7$ s$^{-1}$ and $1.0 \times 10^8$ s$^{-1}$ as well as with PBCs in all directions. After feature extraction, we select atoms of ~10 - 20 Å away from the surfaces or deformation ends to construct the ML datasets.

We also note many glass deformation papers[43,44] employ a strategy of quenching a small cell and replicating it to build a large simulation cell for deformation simulation. This strategy can save a large amount of time and seems to have no significant effects on deformation behaviors. However, in generating data for ML, this will generate replicated site environments that create the potential for ML information leakage, overfitting, and overestimation of score. In this work, we simulate large, un-replicated samples to guarantee the non-duplication of site environments for ML.

## Data availability

The datasets used in this work are available in figshare with the DOI of https://doi.org/10.6084/m9.figshare.7941014.v2.

## Code availability

The featurization and ML codes can be publicly found in our open-source packages *amlearn* https://github.com/Qi-max/amlearn and *matminer* https://github.com/hackingmaterials/matminer.

## Acknowledgements

The authors would like to thank L.F. Zhang, J. Ding, E. Ma and Z. Fan for beneficial discussions. AJ acknowledges support from U.S. Department of Energy, Office of Basic Energy Sciences, Early Career Research Program, which intellectually led the effort. QW also acknowledges the support of National Natural Science Foundation of China (51701190). This research used resources of the National Energy Research Scientific Computing Center (NERSC), a U.S. Department of Energy Office of Science User Facility operated under Contract No. DE-AC02-05CH11231, and the National Supercomputing Center in Shenzhen, China.

## Author contributions

QW and AJ developed the plan for this study. QW designed the features, performed the simulations and developed the machine learning models. QW and AJ analyzed the results and wrote the manuscript. Correspondence and requests for materials should be addressed to QW ([qiwang.mse@gmail.com](qiwang.mse@gmail.com)) and AJ ([ajain@lbl.gov](ajain@lbl.gov)).

## Competing interests

The authors declare no competing interests.